\input phyzzx.tex
\tolerance=1000
\voffset=-0.0cm
\hoffset=0.7cm
\sequentialequations
\def\rl{\rightline}

\def\t1{{\tilde 1}}

\def\PRL#1#2#3{Phys. Rev. Lett. {\bf#1} (19#2) #3}

\def\f{\phi}
\def\t{\theta}

\REF{\COS}{S. Perlmutter et al, Astrophys. J. {\bf 483}, 565 (1997), astro-ph/9608192.}
\REF{\RAT}{B. Ratra and P. J. Peebles, Phys. Rev. {\bf D37}, 3406, (1988).}
\REF{\STE}{R. R. Caldwell, R. Dave and P. J. Steinhardt, \PRL {\bf80} (1998) 1582; L.Wang, R. R. Caldwell, J. Ostriker and P. J. Steinhardt,
Astrophys. J. {\bf 530} (2000) 17.}
\REF{\QUI}{P. Binetruy, hep-ph/0005037.}
\REF{\TRA}{L. Wang, P. J. Steinhardt and I. Zlatev, Phys. Rev. {\bf D59} (1999) 123504, astro-ph/9812313; \PRL {\bf 82} (1999) 896, astro-ph/9807002.}
\REF{\DES}{E. Witten, ``Quantum Gravity in De Sitter Space'', Talk at the Strings 2000 Conference, http://theory.tifr.res.in/strings/.}
\REF{\ACC}{S. Hellerman, N. Kaloper and L. Susskind, hep-th/0104180; W. Fischler, A. Kashani-Poor, R. Mcnees and S. Paban, hep-th/0104181; X. He, astro-ph/0105005.}
\REF{\DIM}{N. Arkani-Hamed, S. Dimopoulos and G. Dvali,  Phys. Lett. {\bf B429} (1998) 263, hep-ph/9803315;
I.~Antoniadis, N.~Arkani-Hamed, S.~Dimopoulos and G.~Dvali,
Phys. Lett. {\bf B436}, 257 (1998), hep-ph/9804398.}
\REF{\BIN}{P. Binetruy, Phys. Rev. {\bf D60} (1999) 063502, hep-th/9810553.}
\REF{\NEW}{J. M. Moffat, hep-th/0105017; C. Deffayet, G. Dvali and G. Gabadadze, astro-pth/0105068.}
\REF{\MAE}{K. Maeda, astro-ph/0012313.}
\REF{\BRO}{J. H. Brodie, hep-th/0101115 and references therein.}
\REF{\AFIV}{G. Aldazabal, A. Font, L. Ibanez and G. Violero, hep-th/9804026; L. Ibanez,  C. Munoz and S. Rigolin, hep-ph/9812397.}
\REF{\INF}{P. J. Peebles and A. Vilenkin, Phys. Rev. {\bf D59} (1999) 063505; astro-ph/9810509.}
\REF{\KUT}{D. Kutasov and A. Giveon, Rev. Mod. Phys. {\bf 71} (1999) 983, hep-th/9802067.}
\REF{\HBD}{E. Halyo, {\it Phys. Lett.} {\bf B387}  (1996) 43, hep-ph/9606423; P. Binetruy and G. Dvali, {\it Phys. Lett.} {\bf B388}  (1996) 241, hep-ph/9606342;
E. Halyo, Phys. Lett. {\bf B454} (1999) 223; hep-ph/9901302.}
\REF{\EH}{E. Halyo, Phys. Lett. {\bf B461} (1999) 109, hep-ph/9905244; JHEP 9909 (1999) 012, hep-ph/9907223.}

\singlespace
\rl{SU-ITP-01-28}
\rl{hep-ph/0105216}
\rl{\today}
\pagenumber=0
\normalspace
\medskip
\bigskip
\titlestyle{\bf{ Hybrid Quintessence with an End or Quintessence from Branes and Large Dimensions}}
\smallskip
\author{ Edi Halyo{\footnote*{e--mail address: vhalyo@stanford.edu}}}
\smallskip
\centerline {Department of Physics}
\centerline{Stanford University}
\centerline {Stanford, CA 94305}
\centerline{and}
 \centerline{California Institute for Physics and Astrophysics}
\centerline{366 Cambridge St.}
\centerline{Palo Alto, CA 94306}
\smallskip
\vskip 2 cm
\titlestyle{\bf ABSTRACT}

We describe a model of hybrid quintessence in which in addition to the tracker field there is a trigger field which is responsible for ending
quintessence. As a result, hybrid quintessence does not suffer from the problems associated with the eternal acceleration of the universe.
We derive the hybrid quintessence potential on branes from the interbrane interactions in string theory and show that it
requires TeV scale strings and two millimeter size dimensions. This scenario predicts a dark energy density of $O(mm^{-4})$ and relates the
smallness of this energy to the large size of the extra dimensions.

\singlespace
\vskip 0.5cm
\endpage
\normalspace

\centerline{\bf 1. Introduction}
\medskip

Recent observations strongly suggest that the expansion of the universe is accelerating and that most of energy density in the universe
is not in the form of baryonic or dark matter[\COS]. The only possible explanation of these results is the existence of some form of dark energy which
has negative pressure. One candidate for the missing energy is the cosmological constant $\Lambda$, i.e. nonzero vacuum energy with $p_{\Lambda}=-
\rho_{\Lambda}$.
Lately, the possibility of another candidate called quintessence has been raised[\RAT,\STE,\QUI]. This is based on an extremely slowly varying and light
scalar field
with an equation of state $w=p/\rho$ in the range $-1<w<-1/3$. If $-1<w<-1/3$, Einstein's equations for a
Robertson--Walker universe shows that the expansion of the universe accelerates. A particular kind of quintessence called a tracker field
has the additional advantage of being very insensitive to initial conditions[\TRA]. Thus, quintessence seems to be an elegant alternative to
a cosmological constant. Unfortunately, there is no dynamical microscopic derivation of neither the quintessence potential nor the very small scale
(compared to the Planck scale) that is needed to explain the smallness of the dark energy density.

On the other hand, the only candidate for a consistent quantum theory of gravity is superstring (or M) theory. The only known formulation of string
theory is in terms of S--matrices which require infinitely separated, free in and out states. This is not a problem in a flat and infinite
universe, i.e. Minkowski space. However, if the universe is dominated by dark energy the situation is quite different. For example, if
there is a (positive) cosmological constant, the universe will asymptotically become a de--Sitter space. In de--Sitter space there is a horizon
(given by the
Hubble constant $H$) and therefore physics is confined to a finite region. As a result, it is impossible to define asymptotic in and out states
and therefore S--matrices. Moreover, the de--Sitter vacuum has a finite temperature which makes the formulation of an S--matrix even more
difficult. Thus, a cosmological constant poses very serious problems for string theory at least as it is presently understood and
formulated[\DES]. Very recently, it was shown that quintessence is also plagued by the same problems mentioned above[\ACC]. The simplest way out of these
problems is for quitessence to end in the (remote) future. In this case, there is neither a future horizon and nor any of the problems
associated with it.
The same is not possible for the cosmological constant since it is not a dynamical field and cannot be turned off.

In this letter, we consider quintessence with two fields which we call hybrid quintessence (due to its conceptual similarity to hybrid inflation). We
show that a tracker potential where the coupling is given by another slowly varying scalar field (which we call the trigger field) results in a
quintessence scenario which comes to an end. In the early universe, when the background energy density (either matter or radiation) dominates both fields are
virtually constant. After some time, the energy density becomes dominated by quintessence as in the usual models. During the
quintessence era, both fields have values of $O(M_P)$ and they roll down their potential very slowly. Eventually, at some later time, the trigger field starts to roll
down fast to the minimum of its potential with a vanishing VEV. When it reaches the minimum it starts to perform damped oscillations about it, i.e. it behaves like
nonrelativistic matter and finally settles at the minimum.
As a result, the whole potential vanishes ending the quintessence era. Most of the original potential energy density gets diluted (as $\sim a(t)^{-3}$) due the expansion
of the universe (i.e.
the damping part of the oscillations). A finite part remains as
the kinetic energy of the tracker field which falls as $\sim a(t)^{-6}$. This energy density becomes negligible after some time
and the universe becomes matter dominated again. The end of quintessence and the succeeding era of matter domination result in an infinite
horizon size, i.e. there is no future horizon. In such a scenario, there are no problems in formulating string theory.

We also show that hybrid quintessence is a generic phenomenon on branes. This in itself is important since the work of the last few
years showed that we may indeed be living on a brane[\DIM]. We derive the quintessence potential on the brane from string theory, i.e. from the brane--brane
interactions mediated by open strings stretched between the branes.
The simplest brane configuration which results in quintessence is given by two widely
separated D3 branes which are at a relative angle. The nonzero relative angle between the D3 branes breaks supersymmetry and
generates a quintessence (tracker) potential with two fields. The tracker field is given by the scalar that parametrizes the distance between
the branes whereas the trigger field parametrizes the relative angle between the branes. We find that the smallness
of the quintessence energy can be explained by the existence of TeV scale strings with two large (mm size) dimensions transverse to the branes.
In this scenario, the smallness of the dark energy density is a direct consequence of the large compactification volume needed for explaining the
hierarchy between the TeV and Planck scales.

This letter is organized as follows. In section 2 we review quintessence and its problems related to future horizons. In section 3, we consider
hybrid quintessence and show that it comes to an end. In section 4
we derive the hybrid quintessence potential on branes using string theory. Section 5 contains our conclusions and a discussion of our results.

\bigskip
\centerline{\bf 2. Quintessence and Future Horizons}
\medskip

Using the Einstein equations for a Robertson--Walker universe with the metric
$$ds^2=-dt^2+a^2(t)(dr^2+r^2 d\Omega^2) \eqno(1)$$
the acceleration of the expansion of the universe is given by
$$3{{\ddot a} \over a}=-4\pi(\rho+3p) \eqno(2)$$
where $\rho$ and $p$ are the energy density and pressure of a perfect fluid filling the universe. Thus, the universe can accelerate only if
the fluid has a negative pressure with $p<-\rho/3$. One such possibility is a nonzero cosmological
constant (or vacuum energy). In this case the equation of state is $w_{\Lambda}=p_{\Lambda}/\rho_{\Lambda}=-1$. On the other hand, the bound
on $w$ from
supernovae observations is $-1<w<-2/3$ so the missing energy does not have to be in the form of a cosmological constant.
Another possibility which has generated a lot of interest lately is quintessence[\RAT, \STE, \QUI]. In this scenario, the missing energy is given by the potential
energy of an extremely light and slowly varying scalar field. For a scalar field $\phi$ with a potential $V(\phi)$ the equation of state is
$$w_{\phi}={{\dot \phi}^2-2V(\phi) \over{{\dot \phi}^2-2V(\phi)}} \eqno(3)$$
When the kinetic energy is less than the potential energy we get negative pressure and $-1<w<0$; when $\dot{\phi}^2<V(\phi)$, $-1<w<-1/3$  and
the universe accelerates. For a very
slowly varying field $w \sim -1$ and quintessence mimics a cosmological constant.

The quintessence potential $V(\phi)$ can take a few generic forms one of which is
$$V(\phi)={\lambda M^{4+n} \over \phi^n} \eqno(4)$$
where $\lambda$ is a coupling constant of $O(1)$ and
$M$ is a small mass scale which depends on $n$ and is fixed by the value of the dark energy. There is no microscopic, dynamical derivation
of this potential and the small scale $M$ to date. (However, for an attempt see [\BIN].)
For this potential there are tracker solutions; i.e. for a very wide range  of initial
conditions the late time evolution of the tracker field $\f$ is the same. Therefore, late cosmology is not sensitive to initial conditions and
quintessence is a very generic phenomenon. It can be shown that at late times when the universe is in the quintessence regime with matter (with $w=0$)
as background
$$\rho \sim a^{-3(1+w)} \eqno(5)$$
and
$$a(t) \sim t^{2 / {3(1+w)}} \eqno(6)$$
where
$$w=-1+{n \over {2+n}} \eqno(7)$$

However, the above behavior of $a(t)$ is problematic if the universe accelerates indefinitely. In such a universe there is a future
event horizon. For the Robertson--Walker metric in eq. (1)
the longest distance a ray of light emitted at time $t_0$ can travel is given by
$$R=\int^{\infty}_{t_0}{dt \over a(t)} \eqno(8)$$
We see that for $a(t)$ in eq. (5) $R$ is finite. Therefore, observers who are separated by a distance larger than $R$ at $t_0$ can never
communicate. Such a universe is very similar to the de--Sitter space; both have horizons and vacua with finite temperature.

Recently, it was argued that the above properties of an eternally accelerating universe, i.e. the future horizon and finite temperature pose
very serious problems for string theory[\ACC]. String theory is formulated as an S--matrix theory which requires infinitely separated
noninteracting in and out states. Due to the horizon, an accelerating universe behaves like a box of finite volume in which there are no
isolated states. Finite temperature of such a universe makes the formulation of an S--matrix even more difficult. Thus, it seems that eternal
quintessence poses the same problems for string theory as a cosmological constant.
The simplest and most conservative solution to these problems is to modify  quintessence so that it ends at a finite time in the (remote) future. (For other
recent solutions see [\NEW].)

\bigskip
\centerline{\bf 3. Hybrid Quintessence}
\medskip

In this section we consider (hybrid) quintessence with a tracker potential of the form in eq. (3) where the coupling $\lambda$ is given by another
scalar field, i.e. $\lambda=\t/M_P$. The potential becomes
$$V(\f,\t)={\t^2 M^{4+n} \over M_P^2 \phi^n} \eqno(9)$$
$\f$ (and $\t$) satisfy the equation of motion
$${\ddot{\f}}+ 3H {\dot \f}+ V^{\prime}(\f)=0 \eqno(10)$$
where $H$ is the the Hubble constant.
The masses for  $\t$ and $\f$ are given by
$$m_{\t}^2 \sim { M^{4+n} \over M_P^2 \phi^n} \eqno(11)$$
and
$$m_{\f}^2 \sim {\t^2 M^{4+n} \over M_P^2 \phi^{n+2}} \eqno(12)$$
The above potential has a tracker solution to which a very large range of initial conditions converge and which satisfies[\RAT, \TRA]
$$V^{\prime \prime} \sim (1-w_{\f}^2)((n+1)/n) H^2 \eqno(13)$$
Thus, as the universe evolves the potential $V$ tracks the decreasing Hubble constant.

In this case, the evolution of the early universe is identical to the usual quintessence models. After inflation and reheating, assuming
equipartition of energy, the universe is dominated by the background (radiation or matter) energy density, $\rho_b>> \rho_{\f}$.
Therefore, the trigger field, $\t$ satisfies the slow--roll condition $H^2>>m_{\t}^2$ and remains constant with a VEV of
$O(M_P)$. On the other hand, the behavior of $\f$ depends on $\rho_{\f}$. Assuming equipartition of energy, $\rho_{\f}$ has a value larger than
$\rho_{tr}$, the energy density corresponding to the tracker solution[\TRA]. In this case, $\f$ rolls down very fast and comes to a stop at a VEV of $O(M_P)$
due to the large redshift of the kinetic energy. At this point $\rho_{\f}<<\rho_{tr}$ and $\t$ and $\f$ are frozen until eq. (13) is satisfied.
The quintessence equation of state is $w_{\f} \sim -1$ since the scalars have very little kinetic energy.

During the radiation or matter dominated era the quintessence energy density changes as $\rho_{\f} \sim a(t)^{-3(1+w_{\f})}$ where
$$w_{\f} \sim {n w_b-2 \over {n+2}} \eqno(14)$$
For matter domination, $w_b=0$ and we get $\rho_{\f} \sim a(t)^{-3n/(n+2)}$. We find that $\rho_{\f}/\rho_b$
increases with time and at some point quintessence starts to dominate the universe.

In the quintessence dominated era, $\f$ is still slowly varying since $H^2 >> m_{\f}^2$ for $\f>M_P$. On the other hand,
$\t$ starts to roll down its potential because $H^2 \sim m^2_{\t}$ for $\t \sim M_P$. The potential starts to decrease slightly due to the
rolling of $\f$ (to larger values) but much more strongly due to the rolling of $\t$ (to smaller values).
The time quintessence dominates the universe is given
by the time it takes for the potential to decrease to $V \sim \dot{\f}^2$. At that point, $w_{\f} \sim -1/3$ and the universe stops accelerating
and starts decelerating. In the meantime, $\t$ continues to to roll down to its minimum at $\t =0$.
The time that this takes is fixed by the VEV of $\t \sim M_P$ at the quintessence era.
Eventually $\t$ reaches its minimum at $t=0$, starts to perform damped oscillations around it and stops at $\t=0$. As a result, the potential vanishes. Most of
the original quintessence energy density decays like nonrelativistic matter due to the expansion of the universe. However, a finite part
remains as the kinetic energy of $\f$ which redshifts much faster than matter as $a(t)^{-6}$. Finally the universe enters a matter dominated
era again.

The end of quintessence solves the problems asscociated with future horizons by eliminating horizons. This can be seen from eq. (8) for the
horizon size. When the universe accelerates from time $t_0$ to $t_1$ due to quintessence with potential given by eq. (9) and $w_{\f} \sim -1$
the horizon size is finite since $t_1$ is finite. When deceleration sets in the horizon size starts to grow and eventually becomes
infinite since
$$R=\int^{\infty}_{t_1}  t^{-2/3}dt \eqno(15)$$
diverges.

\bigskip
\centerline{\bf 4. Hybrid Quintessence on Branes}
\medskip

As mentioned above, there is no microscopic derivation of the quintessence potential or the small scale $M$ to date. A partially successful attempt appears in [\BIN]
where it was shown that the tracker potential in eq. (3) can be obtained from nonperturbative dynamical effects in certain supersymmetric gauge theories
(with no explanation for the very small scale required). However that scenario cannot be
generalized to the two field hybrid quintessence given by  eq. (9) and suffers from the problems related to future horizons. Even if the gauge coupling constant
is made dynamical, e.g. as in a string model
with the dilaton, one gets two fields with runaway behavior rather than only one as above. In the following, we will see that hybrid quintessence
can be generically obtained on branes. We will derive the hybrid quintessence potential from the brane--brane potential in string theory. Previously quintessence on branes
has been considered in the context of [\BIN] in [\MAE] without deriving the potential. It was shown that, the early evolution of quintessence on branes is different due to
new terms in the Friedmann equation. However, these do not affect the late time behavior of quintessence and therefore will not change our results.

The prototypical brane configuration we will consider is given by a D brane on top of an orientifold plane with another D brane at an angle
and at a large distance. The orientifold which has negative tension is needed to cancel the positive brane tension (i.e. to give vanishing cosmological constant on the brane)
and the second brane
is at an angle in order to break supersymmetry and generate the quintessence potential. We will assume that the space transverse to the branes $T^{9-p}$ is orbifolded
so that on the brane there is only $N=1$ supersymmetry before it is broken by the relative angle.
Consider a $Dp$ and a $Dp^{\prime}$ brane parallel and separated by a distance $r$. (We will ignore the orientifold plane, $Op$ in the following since
all it does is to cancel the constant term in the potential.) We will compute the potential on the $Dp-Op$ pair created by the presence
of the $Dp^{\prime}$ brane by using supergravity[\BRO]. This is justified because supergravity is a very good approximation to the full string
theory when the branes are very far from each other (compared to the string scale).

The metric generated by the $Dp^{\prime}$ brane is given by
$$ds^2=h(r)^{-1/2}dx^2_{par}+h(r)^{1/2}dx^2_{perp} \eqno(16)$$
where the subscripts $par$ and $perp$ denote the directions parallel and perpendicular to the brane world--volume and
$$h(r)=1+g_s({1 \over {M_s r}})^{7-p^{\prime}} \eqno(17)$$
with $M_s$ the string scale and $g_s$ the asymptotic value of the string coupling.
The dilaton solution for the brane is
$$e^{-2D}=h(r)^{(p^{\prime}-3)/2} \eqno(18)$$

The above supergravity background gives us the potential on the $Dp$ brane when substituted into the brane action
$$S_p={M_s^{p+1} \over g_s}   \int d^{p+1}y e^{-D} \sqrt{det g_{\mu \nu}} +{M_s^{p+1} \over g_s} \int d^{p+1}y A_{p+1}   \eqno(19)$$
where $g_{\mu \nu}$ is the pullback of the metric in eq. (16) to the $p+1$ dimensional world--volume. $A_{p+1}$ is the Ramond--Ramond potential and the last term
above is only present for the case $p=p^{\prime}$.
Using eqs. (16-19) we get (for $p=p^{\prime}$)
$$V_p(r)={M_s^{p+1} \over g_s} [h(r)^{(p^{\prime}-3)/4} h(r)^{-(p+1)/4}+ h(r)](1+O((\partial X)^2)+ \ldots  \eqno(20)$$
In the expansion of the potential the first term is the $Dp$ brane tension which is cancelled by the negative and equal orientifold tension.
The next terms in the expansion are
$$V_p(r)=M_s^{p+1} \left ({{p^{\prime}-p}\over 4} \right) ({1 \over {M_s r}})^{7-p^{\prime}} +\ldots)  \eqno(21)$$
For $p \not = p^{\prime}$ the last term in eq. (19) is absent and therefore the term $(p^{\prime}-p)/4$ above is replaced with $(p^{\prime}-p-4)/4$. Note that for parallel
branes the potential vanishes either for $p^{\prime}=p$ (when the gravitational and dilatonic potentials are cancelled by the Ramond--Ramond potential) or for
$p^{\prime}=p+4$ (when the gravitational potential cancels the dilatonic one).

For $p=p^{\prime}$, consider a configuration in which one of the branes is rotated relative to the other by $\pi/2$ in one direction (plane) and by $\pi/2+\t$ in another
plane. In this case there is a repulsive potential between the branes. The full potential in eq. (20) becomes[\BRO]
$$V_p(r)= {M_s^{p+1} \over g_s} h(r)^{(p^{\prime}-p-1)/4} (sin^2 \t h(r)^{1/2}+cos^2 \t h(r)^{-1/2})^{1/2} \eqno(22)$$
(For a simple relative angle $\t$ the power of $h(r)$ is replaced by $(p^{\prime}-p+1)$.)
There are two cases where the above potential vanishes because supersymmetry is restored.
First, for $p=p^{\prime}$ the potential vanishes when the branes are rotated by two different angles each equal to $\pi/2$ as above.
Second, the potential vanishes also for $p^{\prime}-p=2$ with only one angle $\t=\pi/2$. This is the configuration with
a $D(p+2)$ brane perpendicular to a $Dp$ brane and is equivalent to the one we are considering.
Including the negative orientifold tension, for small $\t$ and large distances $r>>M_s^{-1}$ the potential in eq. (22) becomes (for $p=p^{\prime}$)
$$V_p(r) \sim M_s^{p+1}(\t^2({1 \over {M_s r}})^{7-p^{\prime}}+ \ldots) \eqno(23)$$
The distance between the branes is parametrized by the scalar field $\f=M_s^2 r$ whereas the angle is parametrized by $\t=M_P^2 \t$
(with an abuse of notation we
denote both the angle and the field by $\t$).

We see that the potential in eq. (23) is exactly of the hybrid quintessence type in eq. (9) for generic values of $p=p^{\prime}$.
The small scale $M$ in the hybrid quintessence potential is given by the string scale $M_s$.
We conclude that hybrid quintessence can be generically obtained on branes.
Consider the case with two D3 branes at an angleas above, i.e. $p=p^{\prime}=3$.
From eq. (23) we get
$$V_3(\f,\t)\sim {\t^2 M_s^8 \over {M_P^2 \f^4}} \eqno(24)$$
In order to obtain the correct vacuum energy, $\rho \sim 10^{-120}M_P^4$ with $\t \sim M_P$, $\f \sim M_P$, we need $M_s \sim 10^{-15}M_P \sim
TeV$. Therefore the string scale has to be $O(TeV)$ if we live on D3 branes with hybrid quintessence. In addition,
since we need $\f \sim M_s^2 r \sim M_P$ we find that the distance between
the branes is $r \sim mm$. Thus the two D5 branes must be at the opposite ends of a mm size large dimension[\DIM]. We find that among the TeV scale
string scenarios the one with two large (mm size) dimensions is preferred by hybrid quintessence models. In this scenario, the smallness
of the dark energy
is directly related to the smallness of the string scale compared to the Planck scale (or the large size of the extra dimensions). For example,
since for two large dimensions $M_P^2 \sim M_s^4 R^2$ we find that during quintessence the potential energy is
$$V(\f,\t) \sim {\t^2 M_s^8 \over M_P^2 \phi^4} \sim {1 \over R^4} \sim mm^{-4}   \eqno(25)$$
More generally, for any scenario with TeV scale strings one obtains (with $\f \sim \t \sim M_P$)
$$V(\f,\t) \sim {1 \over {M_P^2 V_6}} \eqno(26)$$
where $V_6$ is the volume of the compact $T^6$ transverse to the $D3$ branes. The smallness of dark energy, i.e. $V \sim 10^{-120}M_P^4$ compared to the
Planck scale is a direct consequence of the very large compactification volume $V_6$ that is required to generate the hierarchy between $M_P$ and the
TeV scale.

We have obtained hybrid quintessence in the simplest of brane configurations; however we expect our results to hold for other more complicated
but more realistic brane models[\AFIV]. From our discussion about the three cases in which the potential vanishes, it is clear that there are other
potential configurations such as a pair of almost perpendicular $D6-D4$ branes which can result in hybrid quintessence.
In this case, even though hybrid quintessence
can be realized, there are problems with constraints coming from particle physics. For example, if we live on the D4 brane the potential is
$$V_4(\f,\t) \sim {\t^2 M_s^5 \over {M_P^2 \f}} \eqno(27) $$
In order to get the correct magnitude of dark energy with $\t \sim \f \sim M_P$ we need $M_s \sim 10^{-24}M_P \sim keV$ which is clearly ruled out.
For a configuration of two almost prependicular D5 branes (with $3+1$ dimensions in common) it is easy to show that one needs $M_s \sim 10~MeV$ which is also
ruled out.

Seen from the ten dimensional bulk point of view, the evolution of the universe during the quintessence era is as follows. At the beginning of
quintessence when $\t \sim \f \sim M_P$, the two branes are very far from each other and at a relative angle. As long as quintessence is
dominant the distance and angle will be very slowly varying. As the trigger field, $\t$ starts
to roll down its potential one of the branes starts to rotate. Quintessence and the acceleration
of the universe end when the angle is so small that the potential energy is about the kinetic energy. Eventually the brane oscillates around the parallel
configuration and
stops (i.e. $\t=0$ when both branes become parallel and the potential energy vanishes). However, $\f$ is still slowly increasing due to
its kinetic energy which means that the branes are slowly separating. This kinetic energy will decrease very fast with time compared to matter
energy density and will become negligible. From the bulk point of view, using
eq. (10) we see that the brane will come to a stop due to the friction which arises from the Hubble term.

Above, for simplicity we assumed that the $\t<<1$. However, we need the value of the trigger field $\t \geq M_P$ at the beginning of quintessence which
is contrary to this assumption. Using the full potential in eq. (22) one can show that results very similar to ours are obtained. This is clear from
the bulk point of view. Even if the two branes are at a large relative angle all they can do to minimize their energy is to rotate in order to become parallel. So for
large $\t$ the detailed evolution of the fields is more complicated than the one we obtained above but the general behavior is the same.

\bigskip
\centerline{\bf 5. Conclusions and Discussion}
\medskip

In this letter, we described a model of hybrid quintessence in which in addition to the tracker field there is a trigger field which is responsible for ending
quintessence. The potential required for this scenario is a simple generalization of the usual tracker potential. The main virtue of hybrid quintessence is the
fact that it ends. After quintessence the universe goes back to a matter dominated era and the finite horizon size during quintessence starts to grow (to infinity).
 As a result, problems associated with future horizons such as formulation of string theory in such space--times are absent in this scenario. Thus, it seems that (hybrid)
quintessence is a better alternative to dark energy than a cosmological constant if we want to take string cosmology seriously.

We also showed that hybrid quintessence arises naturally on branes and obtained the tracker potential dynamically from brane--brane interactions.
The setup we described above (with two D3 branes at a distance and an angle) is the simplest one
but we believe that more realistic and complicated brane configurations can also lead to hybrid quintessence. We found that the smallness of dark energy is a result of
the smallness of the string scale compared to the Planck scale. Thus, we are led to brane--world scenarios with TeV scale strings. Moreover, the large value of the tracker
field requires that there be mm size dimensions. This singles out scenarios with TeV scale strings with two large (mm size) dimensions as the most favorable for hybrid
quintessence. For a long time it was known that the amount of dark energy density $\sim mm^{-4}$. Our hybrid quintessence scenario predicts this relation
dynamically. In this scenario, the 120 orders of magnitude difference between the dark energy density and Planck energy density is a consequence of the very large
compactification volume required to explain the hierarchy between the TeV and Planck scales.

An intriguing question is whether inflation and quintessence have a common origin[\INF].
The tracker potential which is responsible for hybrid quintessence arises from two branes at a large distance and at an angle. Since the branes are repelling each other
one would assume that they were close to each other at earlier times. However, when the branes are close to each other the physical description changes because
supergravity is no longer a good approximation to string thory. For small brane separations the description is in terms of a gauge theory in which the
potential is dramatically different. In fact, it is well--known that in this regime the tracker potential used above becomes an anomalous D--term for a $U(1)$ gauge
group[\BRO, \KUT]. This raises the possibility of having D--term inflation on the brane followed by hybrid quintessence. Thus, at early universe the branes are close to each
other and D--term inflation takes place[\HBD, \EH]. After inflation ends the branes continue to separate. When they are widely separated the supergravity potential gives rise to
hybrid quintessence.

\bigskip
\centerline{\bf Acknowledgements}

I would like to thank John Brodie for a very useful discussion.

\vfill

\refout

\end
\bye